\documentclass[aps,prl,showpacs,twocolumn,preprintnumbers]{revtex4-1}
\usepackage{mathrsfs}
\usepackage{amsmath}
\usepackage{amsfonts}
\usepackage{amssymb}
\usepackage{amsthm}
\usepackage{graphicx}
\usepackage{natbib}
\usepackage{color}
\usepackage{hyperref}
\usepackage{bm}
\usepackage[caption=false]{subfig}
\usepackage{verbatim}

\begin{document}
\title{Super-Poissonian shot noise of squeezed-magnon mediated spin transport}
\author{Akashdeep Kamra}
\email{akashdeep.kamra@uni-konstanz.de}
\author{Wolfgang Belzig}
\email{wolfgang.belzig@uni-konstanz.de}
\affiliation{Fachbereich Physik, Universit{\"a}t Konstanz, D-78457 Konstanz, Germany}

\begin{abstract}
 The magnetization of a ferromagnet (F) driven out of equilibrium injects pure spin current into an adjacent conductor (N). Such F$|$N bilayers have become basic building blocks in a wide variety of spin based devices. We evaluate the shot noise of the spin current traversing the F$|$N interface when F is subjected to a coherent microwave drive. We find that the noise spectrum is frequency independent up to the drive frequency, and increases linearly with frequency thereafter. The low frequency noise indicates super-Poissonian spin transfer, which results from quasi-particles with effective spin $\hbar^* = \hbar (1 + \delta)$.  For typical ferromagnetic thin films, $\delta \sim 1$  is related to the dipolar interaction-mediated squeezing of F eigenmodes.
\end{abstract} 




\maketitle


{\it Introduction.} The fluctuations of a macroscopic observable, often called {\em noise}, constitute a fundamental manifestation of the underlying microscopic dynamics. While the thermal equilibrium noise is directly related to the linear response coefficients via the fluctuation-dissipation theorem~\cite{Callen1951}, non-equilibrium shot noise provides novel information not accessible via the observable average~\cite{Landauer1998,Blanter2000,Nazarov2003}. Shot noise has been extremely useful in a wide range of phenomena. The optics community has been exploiting intensity shot noise in, among several phenomena~\cite{Walls2008}, observing non-classical photon states~\cite{Slusher1985}. Charge current shot noise has proven to be an effective probe of many-body effects in electronic systems~\cite{Blanter2000,Nazarov2003}. It has also been employed to ascertain the unconventional quanta of charge transfer in the fractional quantum Hall phase~\cite{Jain1989,Kane1994,Saminadayar1997,Reznikov1999} and superconductor-normal metal hybrids~\cite{Kozhevnikov2000,Jehl2000,Cron2001,Lefloch2003}. Noise has furthermore been proposed as a means to observe quantum spin~\cite{Burkard2000} or mode~\cite{Forgues2015} entanglement in electronic circuits.  

Spin current forms an observable of interest in a wide range of systems, such as topological insulators~\cite{Hasan2010}, triplet superconductors~\cite{Kazushige2001}, magnetic insulators~\cite{Kruglyak2010,Bauer2012} and so on, in which the spin degree of freedom plays an active role. While spin dependent charge current noise has been discussed~\cite{Belzig2004,Guerrero2006,Arakawa2011}, the potential of spin current noise has remained largely untamed. Foros {\it et al.} have considered the applied voltage driven, and thus conduction electrons mediated, spin current shot noise in metallic magnetic nanostructures~\cite{Foros2005}. The recent experimental observations of pure spin current thermal noise~\cite{Kamra2014}, and non-equilibrium spin accumulation driven charge current shot noise~\cite{Arakawa2015}, indicate the feasibility of and bring us closer to exploiting this potential. In semiconductor physics, spin noise spectroscopy has already become an established experimental technique~\cite{Oestreich2005,Mueller2010}. 

\begin{figure}[tp]
\begin{center}
\includegraphics[width=85mm]{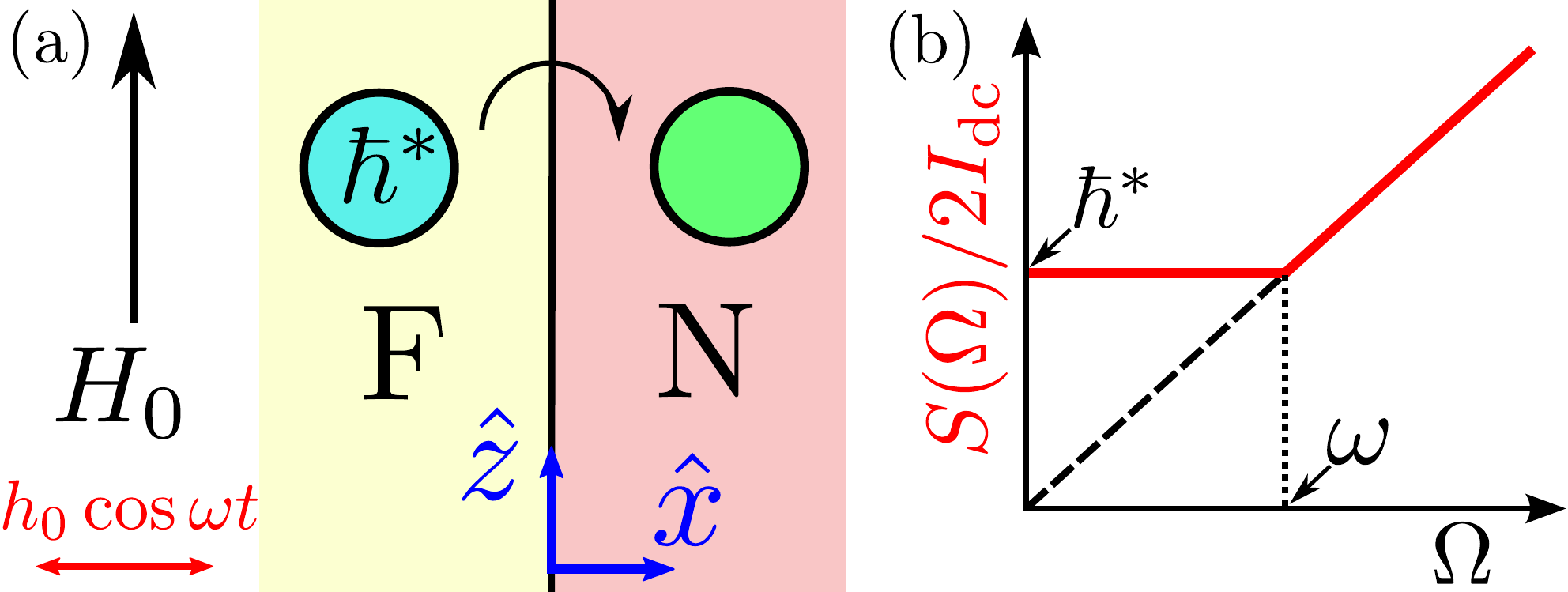}
\caption{(a) Schematic of the ferromagnet (F) and non-magnetic conductor (N) bilayer analyzed in the text. The coordinate system is depicted in blue. A static magnetic field $H_0 \hat{\pmb{z}}$ saturates F magnetization along $\hat{\pmb{z}}$ while a coherent microwave field $h_0 \cos \omega t \ \hat{\pmb{x}}$ creates magnonic excitations in F. The latter annihilate at the interface creating excitations and injecting z-polarized spin current in N. (b) Schematic plot of $S(\Omega)/2I_{\mathrm{dc}}$ vs. $\Omega$ [Eq. (\ref{scnoise})]. $S(\Omega)$ and $I_{\mathrm{dc}}$ are respectively the noise power spectral density and the dc value of the interfacial spin current.} \label{bilayer}
\end{center}
\end{figure}

Heterostructures formed by interfacing a non-magnetic conductor (N) with a ferromagnet (F), typically an insulator (FI), are of particular interest since they allow transfer of pure spin current carried by the collective magnetization dynamics in F to electrons in N. This spin transfer phenomenon has come to be known as {\em spin pumping}~\cite{Tserkovnyak2002}. FI$|$N  bilayers have been the playground for a plethora of newly discovered and proposed effects~\cite{Bauer2012,Weiler2013} making a microscopic understanding of the spin transfer process highly desirable. In this Letter, we investigate spin transfer between the collective magnetization modes in F and electrons in N by examining the zero-temperature spin current shot noise when F is driven by a coherent microwave magnetic field (Fig. \ref{bilayer}). Within the commonly used terminology~\cite{Tserkovnyak2002,Brataas2012}, this may be called {\em coherently driven spin pumping shot noise}. 

The three key findings of this Letter are spontaneous squeezing~\cite{Walls2008} of F eigenmodes, super-Poissonian nature of spin transport, and a non-trivial frequency dependence of the spin current noise power spectral density $S(\Omega)$ [Fig. \ref{bilayer} (b)]:
\begin{align}\label{scnoise}
S(\Omega) = &  \hbar^* \frac{I_{\mathrm{dc}}}{\omega} \left( |\omega + \Omega| + |\omega - \Omega| \right),
\end{align}
with $\omega$ the drive frequency, $I_{\mathrm{dc}}$ the dc spin current, $\hbar^* = \hbar (1 + \delta)$, and the expression for $\delta$ is derived below. If dipolar interaction is disregarded, spin $\hbar$ quasi-particles - magnons~\cite{Holstein1940,Kittel1963}- constitute the collective magnetization eigenmodes in F. Hence, the spin transfer to N is often assumed to take place in lumps of $\hbar$~\cite{Zhang2012,Bender2015,Meier2003}. However, due to the dipolar interaction, the actual F eigenmodes turn out to be squeezed magnon states. Here, the term {\em squeezing} refers to reduction of quantum uncertainty in one quadrature at the expense of increased uncertainty in the other~\cite{Walls2008}. Thus, the super-Poissonian statistic of spin transfer reflects the super-Poissonian distribution~\cite{Walls2008} of the magnon number in the coherent squeezed-magnon state of F generated by the coherent microwave drive. The same shot noise is interpreted in the F eigenbasis as being a result of Poissonian spin transfer via the squeezed-magnon (s-magnon) quasi-particles which have spin $\hbar^*$ [Fig. \ref{bilayer} (a)].


{\it Hamiltonian.} The Hamiltonian for the system of interest, depicted in Fig. \ref{bilayer} (a), comprises of magnetic ($\tilde{\mathcal{H}}_{\mathrm{F}}$), electronic ($\tilde{\mathcal{H}}_{\mathrm{N}}$), interaction between F and N ($\tilde{\mathcal{H}}_{\mathrm{int}}$), and microwave drive ($\tilde{\mathcal{H}}_{\mathrm{drive}}$) contributions:
\begin{align}\label{htot}
\tilde{\mathcal{H}} = \tilde{\mathcal{H}}_{\mathrm{F}} + \tilde{\mathcal{H}}_{\mathrm{N}} + \tilde{\mathcal{H}}_{\mathrm{int}} + \tilde{\mathcal{H}}_{\mathrm{drive}},
\end{align}
where tilde is used to denote operators. We first evaluate $\tilde{\mathcal{H}}_{\mathrm{F}}$ by quantizing the classical magnetic Hamiltonian $\mathcal{H}_{\mathrm{F}}$ which includes contributions from Zeeman, anisotropy, exchange and dipolar interactions~\cite{Akhiezer1968,Kittel1963}: $
\mathcal{H}_{\mathrm{F}}  = \int_{V_\mathrm{F}} d^3 r \left( H_{\mathrm{Z}} + H_{\mathrm{aniso}} + H_{\mathrm{ex}} + H_{\mathrm{dip}}  \right), $ with $V_{\mathrm{F}}$ the volume of the ferromagnet. An applied static magnetic field $H_0 \hat{\pmb{z}}$ saturates the F magnetization $\pmb{M}$ along the z-direction such that $M_{x,y} ( \ll M_z \approx M_s )$ become the field variables describing the excitations. $M_s$ is the saturation magnetization. We retain terms up to second order in $M_{x,y}$. Employing the relation $M_x^2 + M_y^2 + M_z^2 = M_s^2$ and dropping the constant terms, the Zeeman and anisotropy contributions are obtained as~\cite{Kittel1958,Kamra2015}: $H_{\mathrm{Z}} + H_{\mathrm{aniso}}  = (\omega_0/2 |\gamma| M_s) \left( M_x^2 + M_y^2 \right),$ with $\omega_0 = |\gamma|[\mu_0 H_0 + 2(K_1 + K_u)/M_s]$, where $\gamma = - |\gamma|$ is the typically negative gyromagnetic ratio of F, $\mu_0$ is the permeability of free space, $K_u (>0)$ and $K_1 (>0)$ respectively parameterize uniaxial and cubic magnetocrystalline anisotropies~\cite{Chikazumi1997}. The exchange contribution is~\cite{Kittel1963,Kamra2015}: $H_{\mathrm{ex}} = (A/M_s^2) \left[  \left(\pmb{\nabla} M_x \right)^2 + \left( \pmb{\nabla} M_y \right)^2 \right],$ with $A$ the exchange constant~\cite{Kittel1949}. The dipolar interaction is treated within a mean field approximation via the so called demagnetization field $\pmb{H}_m$ produced by the magnetization: $H_{\mathrm{dip}}  = - (1/2) \mu_0 \pmb{H}_m \cdot \pmb{M}$. For spatially constant $\pmb{M}$, $\pmb{H}_m = - ( N_x M_x \hat{\pmb{x}} + N_y M_y \hat{\pmb{y}} + N_z M_z \hat{\pmb{z}})$ with $N_{x,y,z}$ the elements of the demagnetization tensor, which is diagonal in the chosen coordinate system~\cite{Akhiezer1968}.

The classical magnetic Hamiltonian is quantized by defining the magnetization operator $\tilde{\pmb{M}} = - |\gamma| \tilde{\pmb{S}}_{\mathrm{F}}$~\cite{Kittel1963,Akhiezer1968} with $\tilde{\pmb{S}}_{\mathrm{F}}$ the F spin density operator. The magnetization is expressed in terms of Bosonic excitations by the Holstein-Primakoff transformations~\cite{Holstein1940,Kittel1963}: $\tilde{M}_+ = \sqrt{2 |\gamma| \hbar M_s} \ [1 - (|\gamma| \hbar/ 2 M_s) \tilde{a}^\dagger \tilde{a}] \tilde{a}$, $\tilde{M}_- = \sqrt{2 |\gamma| \hbar M_s} \ \tilde{a}^\dagger [1 - (|\gamma| \hbar/ 2 M_s) \tilde{a}^\dagger \tilde{a}] $, and $\tilde{M}_z = M_s - |\gamma|\hbar \tilde{a}^\dagger \tilde{a}$, where $\tilde{M}_{\pm} = \tilde{M}_x \pm i (\gamma/|\gamma|) \tilde{M}_y$. The operator  $\tilde{a}^\dagger \equiv \tilde{a}^\dagger (\pmb{r})$ creates a magnon at position $\pmb{r}$, satisfies the Bosonic commutation relation: $[\tilde{a}(\pmb{r}), \tilde{a}^\dagger (\pmb{r}^\prime)] = \delta(\pmb{r} - \pmb{r}^\prime)$, and is expressed in terms of the Fourier space magnon creation operators $\tilde{b}^\dagger_{\pmb{q}}$ via $\tilde{a}^\dagger(\pmb{r}) =  \sum_{\pmb{q}} \phi_{q}^*(\pmb{r}) \tilde{b}^\dagger_{\pmb{q}}$ with plane wave eigenstates $\phi_{q}(\pmb{r}) = (1/\sqrt{V_{\mathrm{F}}}) \exp(i \pmb{q}\cdot \pmb{r})$. Following the quantization procedure~\cite{Kittel1963,Akhiezer1968}, the magnetic Hamiltonian simplifies to:
\begin{align}\label{hfquant1}
\tilde{\mathcal{H}}_{\mathrm{F}} & = \sum_{\pmb{q}} \left[ A_{\pmb{q}} \tilde{b}^\dagger_{\pmb{q}} \tilde{b}_{\pmb{q}} +  B_{\pmb{q}}^*  \tilde{b}^\dagger_{\pmb{q}} \tilde{b}^\dagger_{-\pmb{q}} + B_{\pmb{q}} \tilde{b}_{\pmb{q}} \tilde{b}_{-\pmb{q}} \right],
\end{align}
where $A_{\pmb{q}} = A_{-\pmb{q}} = \hbar(\omega_0 + D q^2 + |\gamma|M_s \mu_0 (N_{xz} + N_{yz})/2) + \hbar \omega_A(\pmb{q})$ and $B_{\pmb{q}} = B_{-\pmb{q}} = \hbar |\gamma|M_s \mu_0 N_{xy}/4 + \hbar \omega_B(\pmb{q})$. Here, $D = 2 A |\gamma|/M_s$, $N_{xy} = N_x - N_y$ and so on, $\omega_{A,B}(\pmb{q})$ are the dipolar interaction contributions for magnons with $\pmb{q} \neq \pmb{0}$~\cite{Kittel1963,Akhiezer1968}, and $\omega_{B}(\pmb{q})$ is complex in general. The Hamiltonian (\ref{hfquant1}) is diagonalized by a Bogoliubov transformation~\cite{Holstein1940,Kittel1963} to new Bosonic excitations defined by $\tilde{\beta}_{\pmb{q}} = u_{\pmb{q}} \tilde{b}_{\pmb{q}} - v_{\pmb{q}}^* \tilde{b}_{-\pmb{q}}^\dagger$, 
\begin{align}\label{hf}
\tilde{\mathcal{H}}_{\mathrm{F}} & = \sum_{\pmb{q}} \hbar \omega_{\pmb{q}} \tilde{\beta}^\dagger_{\pmb{q}} \tilde{\beta}_{\pmb{q}},
\end{align}
with transformation parameters:  $\hbar \omega_{\pmb{q}} = \sqrt{A_{\pmb{q}}^2 - 4  |B_{\pmb{q}}|^2  }$, $v_{\pmb{q}} = - 2 B_{\pmb{q}}/\sqrt{(A_{\pmb{q}} + \hbar \omega_{\pmb{q}})^2 - 4  |B_{\pmb{q}}|^2} $, $v_{\pmb{q}}/u_{\pmb{q}} = - 2 B_{\pmb{q}}/(A_{\pmb{q}} + \hbar \omega_{\pmb{q}})$, and $u_{\pmb{q}}^2 = 1 + |v_{\pmb{q}}|^2 $. Here, $u_{\pmb{q}}$ has been chosen to be real positive while $v_{\pmb{q}}$ is in general complex, with $v_{\pmb{0}}$ real. 

If the dipolar interaction is disregarded, $B_{\pmb{q}} = 0$, $\tilde{\beta}_{\pmb{q}} = \tilde{b}_{\pmb{q}}$, and magnon modes are the eigenstates of F. To gain insight into the effect of the dipolar interaction on the eigenmodes, we note that the vacuum corresponding to the new excitations $| 0 \rangle_{\beta}$ is defined by $(u_{\pmb{q}} \tilde{b}_{\pmb{q}} - v_{\pmb{q}}^* \tilde{b}^\dagger_{-\pmb{q}} )$ $| 0 \rangle_{\beta} = 0 $. Employing Baker-Hausdorff lemma and relegating detailed derivations to the Supplemental Material~\cite{SupplMat}, this becomes $ \tilde{S}_2(\xi_{\pmb{q}}) \tilde{b}_{\pmb{q}} \tilde{S}_2^\dagger(\xi_{\pmb{q}}) | 0 \rangle_{\beta} = 0$ with $\xi_{\pmb{q}} = - ( v_{\pmb{q}} / |v_{\pmb{q}}| ) \  \tanh^{-1}(|v_{\pmb{q}}|/u_{\pmb{q}}) $, where $\tilde{S}_2(\xi_{\pmb{q}}) = \exp( \xi_{\pmb{q}}^* \tilde{b}_{\pmb{q}} \tilde{b}_{-\pmb{q}} - \xi_{\pmb{q}} \tilde{b}_{\pmb{q}}^\dagger \tilde{b}_{-\pmb{q}}^\dagger )$ is the two-mode squeeze operator~\cite{Walls2008}, considering $\pmb{q} \neq \pmb{0}$. This leads to $| 0 \rangle_{\beta} = \tilde{S}_2(\xi_{\pmb{q}}) | 0 \rangle_{b}$ showing that the $\tilde{\beta}_{\pmb{q}}$ vacuum is obtained by squeezing the magnon vacuum, two modes ($\tilde{b}_{\pm\pmb{q}}$) at a time. In other words, $\beta_{\pmb{q}}$ excitations are obtained by squeezing $\tilde{b}_{\pm\pmb{q}}$, and are thus called {\em squeezed-magnons} (s-magnons). Instead of deriving a similar relation for the $\pmb{q} = \pmb{0}$ mode, we demonstrate its squeezing by evaluating the vacuum fluctuations of $\tilde{\mathcal{M}}_{x,y} = \int_{V_{\mathrm{F}}} \tilde{M}_{x,y} d^3 \pmb{r} \propto (b_{\pmb{0}}^\dagger \pm b_{\pmb{0}})$:
\begin{align}
 \left\langle \left( \delta \tilde{\mathcal{M}}_{x,y} \right)^2 \right\rangle_{0} & = \frac{|\gamma| \hbar \mathcal{M}_0}{2} \exp \left( \mp 2 \xi_{\pmb{0}} \right), 
\end{align}
where $\langle ~ \rangle_0$ denotes expectation value in the ground state, $\mathcal{M}_0 = M_s V_{\mathrm{F}}$ is the total magnetic moment, and $\xi_{\pmb{0}} =  - \tanh^{-1}(v_{\pmb{0}}/u_{\pmb{0}})$ is real. The sign of $\xi_{\pmb{0}}$, and thus the direction (x or y) of squeezing, is determined by the sign of $- v_{\pmb{0}}/u_{\pmb{0}} \propto B_{\pmb{0}} \propto N_{xy}$. Hence we find reduced quantum noise in one component of the total magnetic moment while the noise is increased in the other component. Owing to dipolar interactions, the F ground state exhibits spontaneous squeezing.

The electronic Hamiltonian for N can be written as $\tilde{\mathcal{H}}_{\mathrm{N}} = \sum_{\pmb{k},s = \pm}  \hbar \omega_{\pmb{k}} \tilde{c}^\dagger_{\pmb{k},s} \tilde{c}_{\pmb{k},s} $, where $\tilde{c}^\dagger_{\pmb{k},s}$ are Fermionic operators that create electrons with spin $s \hbar/2$ along the z-direction in orbitals with wave functions $\psi_{\pmb{k}}(\pmb{r})$. We consider that F and N couple via an interfacial exchange interaction parametrized by $\mathcal{J}$~\cite{Zhang2012,Bender2015}:
\begin{align}
\tilde{\mathcal{H}}_{\mathrm{int}} & = - \frac{\mathcal{J}}{\hbar^2} \int_{\mathcal{A}} d^2 \mathcal{\varrho} \left( \tilde{\pmb{S}}_{\mathrm{F}}(\pmb{\varrho}) \cdot \tilde{\pmb{S}}_{\mathrm{N}}(\pmb{\varrho})  \right),
\end{align}
where $\mathcal{A}$ denotes the interfacial area and $\pmb{\varrho}$ is the interfacial 2D position vector. $ \tilde{\pmb{S}}_{\mathrm{N}} = (\hbar/2)  \sum_{s,s^\prime} \tilde{\Psi}^\dagger_s \pmb{\sigma}_{s,s^\prime} \tilde{\Psi}_{s^\prime} $ is the N spin density operator, where $\tilde{\Psi}_s(\pmb{r}) = \sum_{\pmb{k}} \psi_{\pmb{k}}(\pmb{r}) \tilde{c}_{\pmb{k},s}$ annihilates electron with spin $s \hbar/2$ at $\pmb{r}$, and the components of $\pmb{\sigma}$ are the Pauli matrices. In terms of the normal mode operators~\footnote{We have disregarded the electron spin conserving terms in $\tilde{\mathcal{H}}_{\mathrm{int}}$ since they do not contribute to net z-polarized spin transport~\cite{Zhang2012}.},
\begin{align}
\tilde{\mathcal{H}}_{\mathrm{int}} & = \sum_{\pmb{k}_1 \pmb{k}_2 \pmb{q}} \hbar W_{\pmb{k}_1 \pmb{k}_{2} \pmb{q}} \ \tilde{c}^\dagger_{\pmb{k}_{1} + } \tilde{c}_{\pmb{k}_{2} - } \tilde{b}_{\pmb{q}} \ + \ \mathrm{h.c.} \ ,
\end{align}
with $\tilde{b}_{\pmb{q}} =  u_{\pmb{q}} \tilde{\beta}_{\pmb{q}} + v_{\pmb{q}}^* \tilde{\beta}^\dagger_{-\pmb{q}} $, and $ \hbar W_{\pmb{k}_1 \pmb{k}_{2} \pmb{q}} = \mathcal{J} \sqrt{M_s/2 |\gamma| \hbar} \ \int_{\mathcal{A}} d^2 \varrho \  \psi_{\pmb{k}_1}^*(\pmb{\varrho}) \psi_{\pmb{k}_2}(\pmb{\varrho}) \phi_{\pmb{q}}(\pmb{\varrho})$.
The microwave drives the system via Zeeman coupling between its magnetic field $h_0 \cos (\omega t) \hat{\pmb{x}}$ and the F total magnetic moment $\pmb{\mathcal{M}}$: 
\begin{align}
\tilde{\mathcal{H}}_{\mathrm{drive}} & =  - \mu_0 h_0 \cos (\omega t) B \left( \tilde{\beta}_{\pmb{0}} + \tilde{\beta}^{\dagger}_{\pmb{0}} \right),
\end{align}
with $B = (u_{\pmb{0}} + v_{\pmb{0}}) \sqrt{|\gamma|\hbar \mathcal{M}_0/ 2}$.

Since the magnonic excitations possess spin along the z-direction, we are interested in z-polarized spin current injected into N by F. The corresponding spin current operator is given by:
\begin{align}
\tilde{I_z}  & = \frac{1}{i \hbar} [ \tilde{\mathcal{S}}_{z}, \tilde{\mathcal{H}}_{\mathrm{int}} ] =  \sum_{\pmb{k}_1 \pmb{k}_2 \pmb{q}} - i \hbar  W_{\pmb{k}_1 \pmb{k}_{2} \pmb{q}} \ \tilde{c}^\dagger_{\pmb{k}_{1} + } \tilde{c}_{\pmb{k}_{2} - } \tilde{b}_{\pmb{q}}  +  \mathrm{h.c.}  , \nonumber
\end{align}
with $\tilde{\pmb{\mathcal{S}}} = \int_{V_{\mathrm{N}}} d^3 r \tilde{\pmb{S}}_{\mathrm{N}} (\pmb{r}) $, where $V_{\mathrm{N}}$ denotes the volume of N.


{\it Equations of motion.} We have thus expressed the total Hamiltonian and the spin current operator in terms of the creation and annihilation operators of F (s-magnons) and N (electrons) eigenmodes. Working in the Heisenberg picture, the time resolved expectation value of an observable can be obtained by evaluating the time evolution of electron and s-magnon operators. Since the microwave drives the $\pmb{q} = \pmb{0}$ magnetic mode coherently leaving all other modes essentially unperturbed, we make the quasi-classical approximation replacing $\tilde{\beta}_{\pmb{q}}$ by c-numbers $\beta \delta_{\pmb{q},\pmb{0}}$, and derive the dynamical equation for $\beta(t) = \langle \tilde{\beta}_{\pmb{0}} (t) \rangle$ below. This `approximation' is equivalent to disregarding the equilibrium noise and allows us to focus on the shot noise. The contribution of thermal and vacuum noises shall be considered elsewhere. 

The Heisenberg equations of motion $\dot{\tilde{c}}_{\pmb{k} +} = (1/i\hbar) [\tilde{c}_{\pmb{k} +}, \tilde{\mathcal{H}}]$ simplify to:
\begin{align}\label{eomck+}
\dot{\tilde{c}}_{\pmb{k} +} & = - i \omega_{\pmb{k}} \tilde{c}_{\pmb{k} +} - i \sum_{\pmb{k}_2,\pmb{q}} W_{\pmb{k},\pmb{k}_2,\pmb{q}}  \tilde{c}_{\pmb{k}_2 -} \tilde{b}_{\pmb{q}}.
\end{align}
Similarly, equations of motion can be obtained for $\tilde{c}_{\pmb{k} -}$ and $\tilde{\beta}_{\pmb{q}}$. As detailed in the Supplemental Material~\cite{SupplMat}, we obtain solutions to these equations up to the lowest non-vanishing order in $\mathcal{J}$ using the method employed by Gardiner and Collett~\cite{Gardiner1985} in deriving the input-output formalism~\cite{Walls2008}. Until some initial time $t_0$, F and N do not interact with each other and are in equilibrium so that the density matrix of the system, which stays the same in the Heisenberg picture, factors into the equilibrium density matrices of F and N. The terms $\tilde{\mathcal{H}}_{\mathrm{int}}$ and $\tilde{\mathcal{H}}_{\mathrm{drive}}$ are turned on at $t = t_0$. The steady state solution for any time $t > t_0$ is obtained by taking the limit $t_0 \to -\infty$ in the end. The general solution to Eq. (\ref{eomck+}) for $t>t_0$ can then be written as~\cite{Gardiner1985}:
\begin{align}
 \tilde{c}_{\pmb{k} +}(t)  = & e^{- i \omega_{\pmb{k}}(t - t_0)} \tilde{c}_{\pmb{k} +}(t_0)  \\
   & - i  \sum_{\pmb{k}_2,\pmb{q}} W_{\pmb{k},\pmb{k}_2,\pmb{q}}  \int_{t_0}^{t}  e^{- i \omega_{\pmb{k}}(t - t^\prime)} \tilde{c}_{\pmb{k}_2 -}(t^\prime) \tilde{b}_{\pmb{q}}(t^\prime) d t^\prime. \nonumber
\end{align}
Employing analogous expressions for $\tilde{c}_{\pmb{k} -}$, the Heisenberg equation of motion for $\tilde{\beta}_{\pmb{0}}$, and retaining terms up to second order in $\mathcal{J}$, we obtain the dynamical equation for $\beta(t) = \langle \tilde{\beta}_{\pmb{0}} (t) \rangle$:
\begin{align}\label{betadyn}
\dot{\beta}  = & - i \omega_{\pmb{0}} \beta - (u_{\pmb{0}}^2 + v_{\pmb{0}}^2) \Gamma_{\mathrm{N}} \beta + 2 u_{\pmb{0}} v_{\pmb{0}} \Gamma_{\mathrm{N}} \beta^* \nonumber \\
 &   + i \frac{\mu_0 h_0 B}{\hbar} \cos (\omega t),
\end{align}
where $\Gamma_{\mathrm{N}} = \omega \alpha^\prime = \omega \pi |W_{\epsilon_{\mathrm{Fermi}},\pmb{0}}|^2 V_{\mathrm{N}}^2 \hbar^2 g^2(\epsilon_{\mathrm{Fermi}})  $ represents the magnetic dissipation caused by the electronic bath in $\mathrm{N}$. Here $g(\epsilon_{\mathrm{Fermi}})$ is the electronic density of states at the Fermi energy $\epsilon_{\mathrm{Fermi}}$, and we assume that $W_{\pmb{k}_1,\pmb{k}_2,\pmb{0}} = W_{\epsilon_{\mathrm{Fermi}},\pmb{0}}$ depends only on $\pmb{k}_{1,2}$ magnitudes, and hence on $\epsilon_{\mathrm{Fermi}}$. So far, we have not considered any internal dissipation in F. This can be done by including non-linear interactions with another bath (electrons, phonons, (s-)magnons etc.) in $\tilde{\mathcal{H}}_{\mathrm{F}}$~\cite{Gardiner1985}. The resulting dynamical equation for $\beta$ is obtained by replacing $\Gamma_{\mathrm{N}}$ by $\Gamma = \Gamma_{\mathrm{F}} + \Gamma_{\mathrm{N}}$ in Eq. (\ref{betadyn}), where $\Gamma_{\mathrm{F}}$ depends on the details of the non-linear interaction considered in $\tilde{\mathcal{H}}_{\mathrm{F}}$.


{\it Results and Discussion.} Substituting the ansatz $\beta = \beta_+ \exp(i \omega t) + \beta_- \exp(- i \omega t)$ in Eq. (\ref{betadyn}), we find that $\beta_+ \ll \beta_-$ for $\Gamma \ll \omega_{\pmb{0}}$, and hence $\beta_+$ is disregarded making the rotating wave approximation:
\begin{align}\label{betares}
\beta(t) = & \frac{\mu_0 h_0 B}{2 \hbar} \ \frac{1}{(\omega_{\pmb{0}} - \omega) - i \Gamma (u_{\pmb{0}}^2 + v_{\pmb{0}}^2)} \ e^{- i \omega t}.
\end{align}
Thus we obtain resonant excitation of the $\pmb{q} = \pmb{0}$ s-magnon mode at $\omega = \omega_{\pmb{0}}$. The analysis leading to Eq. (\ref{betadyn}) is employed to obtain the expectation value of the spin current operator up to the order $\mathcal{J}^2$:
\begin{align}
I_z(t) = & \langle \tilde{I}_z(t) \rangle  =  I_{\mathrm{dc}} =  \ 2 \hbar \alpha^\prime \omega |\beta|^2. \label{sc1}
\end{align} 
Thus the spin current injection also exhibits resonant behavior akin to magnetization dynamics~\footnote{We note that our results for magnetization dynamics [Eq. (\ref{betares})] and spin current injection [Eq. (\ref{sc1})] are identical to those obtained by a Landau-Lifshitz-Gilbert (LLG) equation~\cite{Akhiezer1968} plus spin pumping~\cite{Tserkovnyak2002} approach, provided the phenomenological parameters of the latter approach are appropriately identified in terms of our microscopic parameters.}.

The {\em single-sided} spectral density of spin current noise $S(\Omega)$ is obtained via the Wiener-Khintchine theorem for non-stationary processes~\cite{Howard2004}: $S(\Omega) =  2 \int_{-\infty}^{\infty} R(t) e^{i \Omega t} dt$ with $R(t) =  \mathrm{lim}_{\tau_0 \to \infty}  (1/2\tau_0) \int_{- \tau_0}^{\tau_0} \Phi(\tau,\tau-t) d \tau$, where $\Phi(t_1,t_2) = (1/2) \langle  \tilde{\delta I}_z (t_1) \tilde{\delta I}_z (t_2) +  \tilde{\delta I}_z (t_2)  \tilde{\delta I}_z (t_1) \rangle$ is the expectation value of the symmetrized spin current fluctuations [$\tilde{\delta I}_z = \tilde{I}_z - \langle \tilde{I}_z \rangle $] correlator. Assuming zero temperature and again retaining terms up to order $\mathcal{J}^2$, the spin current shot noise simplifies to Eq. (\ref{scnoise}) with $\hbar^* = \hbar (1 + 2 v_{\pmb{0}}^2)$, which is the main result of this Letter.

The zero frequency noise thus becomes $S(0) = 2 \hbar (1 + 2 v_{\pmb{0}}^2) I_{\mathrm{dc}}$ [Eq. (\ref{scnoise})]. Equations (\ref{betares}) and (\ref{sc1}) show that $S(0)$ exhibits resonant behavior as a function of $\omega$. Under certain conditions, the low frequency shot noise for a Poissonian transport process with transport quantum $q$ and dc current $I_0$ is known to be $2 q I_0$~\cite{Blanter2000,Walls2008}. Thus, in the N eigenbasis, in which electrons undergo spin flips by absorbing magnons, our result for low frequency spin current shot noise can be understood as due to correlated spin transfer in lumps of $\hbar$. This interpretation is corroborated by the squeeze parameter $\xi_{\pmb{0}}$ dependent super-Poissonian distribution of the particle (in this case, magnon) number in a coherent squeezed state~\cite{Walls2008}.    

An alternate interpretation for the low frequency noise is obtained in the F eigenbasis: spin transport takes place via the coherent state driven Poissonian transfer~\cite{Walls2008} of $\beta_{\pmb{0}}$ s-magnons which have a spin of $\hbar^* = \hbar(1 + \delta)$ with $\delta = 2 v_{\pmb{0}}^2$. This non-integral spin of s-magnons can also be obtained directly by evaluating the expectation value of the z-component of the total spin in F: $\int_{V_{\mathrm{F}}} \langle \tilde{S}_{\mathrm{F}}^z (\pmb{r}) \rangle d^3 r = - \mathcal{M}_0/|\gamma| + \sum_{\pmb{q}} \hbar (1 + 2 |v_{\pmb{q}}|^2) n_{\pmb{q}}^\beta + \sum_{\pmb{q}} \hbar |v_{\pmb{q}}|^2 $, where the last term in this expression represents the vacuum noise~\cite{Holstein1940}, and $n_{\pmb{q}}^\beta$ denotes the number of s-magnons with wavevector $\pmb{q}$. Thus we see that s-magnon with wavevector $\pmb{q}$ has spin $\hbar(1 + 2  |v_{\pmb{q}}|^2 )$. 

However, $v_{\pmb{q}}$ is considerable only when the relative contribution of the dipolar interaction to the total eigenmode energy $\hbar \omega_{\pmb{q}}$ is {\em not} negligible. In particular, with $\omega_0/2 \pi = 1$ GHz, $\delta = 2 v_{\pmb{0}}^2 \approx 0.4$ for yttrium iron garnet ($|\gamma| = 1.8 \times 10^{11}$ Hz/T, $M_s = 1.4 \times 10^5$ A/m~\cite{Chikazumi1997}) and  $\delta \approx 3.0$ for iron ($|\gamma| = 1.8 \times 10^{11}$ Hz/T, $M_s = 1.7 \times 10^6$ A/m~\cite{Chikazumi1997}) thin films ($N_x = 1, N_{y,z} = 0$). $\delta (\propto N_{xy}^2 )$ vanishes when $N_{xy} = 0$, and $\delta \to 0$ when $H_0/M_s \to \infty$.

\begin{figure}[tb]
\begin{center}
\includegraphics[width=85mm]{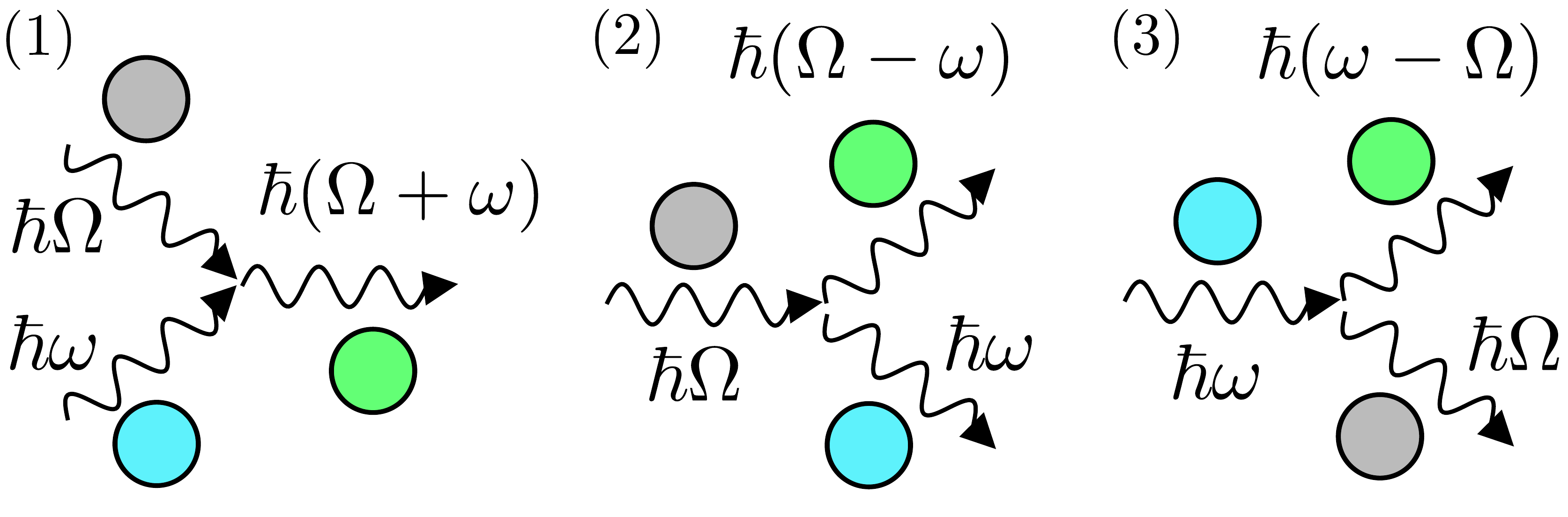}
\caption{Processes contributing to spin current noise at frequency $\Omega$. The blue, green and grey circles respectively depict s-magnon, excitation created in N, and spin current analog of a photon (see text). For $\Omega < \omega$ (the drive frequency), only processes (1) and (3) are allowed. While for $\Omega > \omega$, only processes (1) and (2) take place.} \label{feyndias}
\end{center}
\end{figure}

To discuss a physical understanding of the spin current shot noise frequency dependence [Eq. (\ref{scnoise})], we note that the charge current noise at frequency $\Omega$ is due to absorption and emission of photons at the same frequency~\cite{Nyquist1928}. We make an analogous interpretation of spin current noise in terms of absorption and emission of photon-like quasi particles, keeping in mind that the analogy is mathematical. Thus, for $\Omega < \omega$, the only possible processes are absorption of photon-like quasi-particle and s-magnon while creating an excitation in N [Process (1) in Fig. \ref{feyndias}], and absorption of s-magnon while creating a photon-like quasi-particle and an excitation in N [Process (3) in Fig. \ref{feyndias}]. The rate of each process is proportional to the number of states available for creating an excitation in N, which, at zero temperature, is proportional to the energy of the N excitation governed by energy conservation in the process. Similar arguments can be made when $\Omega > \omega$ (Fig. \ref{feyndias}) thereby motivating the frequency dependence in Eq. (\ref{scnoise}).


{\it Summary.} We have evaluated the zero temperature shot noise of spin current injected into a non-magnetic conductor (N) by an adjacent ferromagnet (F) driven by a coherent microwave drive. The low frequency shot noise indicates spin transfer in quanta of $\hbar^* = \hbar (1 + \delta)$ associated with the zero wavevector excitations in F. We demonstrate that owing to dipolar interaction~\footnote{Mathematically, any bilinear term in $\tilde{\mathcal{H}}_{\mathrm{F}}$ that breaks the axial symmetry about the equilibrium magnetization direction leads to squeezing of the F ground state and excitations}, the F ground state exhibits spontaneous squeezing~\cite{Walls2008}, and its normal excitations are squeezed-magnons with non-integral spin. Our work thus provides important new insights into the magnetization mediated spin transfer mechanism in F$|$N bilayers, and paves the way for exploiting the spontaneously squeezed F ground state.

We gratefully acknowledge valuable discussions with S. T. B. Goennenwein, H. Huebl, R. Gross, Y. M. Blanter, G. E. W. Bauer, and B. Hillebrands. We acknowledge financial support from the DFG through SFB 767 and the Alexander von Humboldt Foundation.

\bibliography{SC_shot_noise}


\widetext
\clearpage
\setcounter{equation}{0}
\setcounter{figure}{0}
\setcounter{table}{0}
\makeatletter
\renewcommand{\theequation}{S\arabic{equation}}

\begin{center}
\textbf{\large Supplemental Material with the manuscript Super-Poissonian shot noise of squeezed-magnon mediated spin transport by} \\
\vspace{0.3cm}
Akashdeep Kamra and Wolfgang Belzig
\vspace{0.2cm}
\end{center}

\setcounter{page}{1}

\section{Squeezing of magnons}

Here we demonstrate that the Bogoliubov transformation required to diagonalize the magnetic Hamiltonian (Eq. (3) in the main text) expressed in terms of the magnon operators results in eigenmodes obtained by squeezing of the magnon modes which are thus called squeezed-magnons (s-magnons). To this end, we first discuss the definitions~\cite{Walls2008,Gerry2004} of the squeeze operator and the squeezed vacuum which will allow us to obtain the desired mathematical relation between the two kinds of excitations. 

For a single mode represented by the annihilation operator $\tilde{a}$, the squeeze operator is defined as:
\begin{align}
\tilde{S}(\xi) = & \exp \left[ \frac{1}{2} \left(  \xi^* \tilde{a}^2 - \xi \tilde{a}^{\dagger 2} \right)   \right]  ,
\end{align}
with $\xi = r \exp (i \theta)$, where $r$ is known as the squeeze parameter and $\theta$ specifies the direction of squeezing~\cite{Gerry2004}. One may thus define a new ``squeezed'' state $| \psi \rangle_s$:
\begin{align}
| \psi \rangle_s = & \tilde{S}(\xi) | \psi \rangle,
\end{align}
in terms of the original state $| \psi \rangle$. When $| \psi \rangle$ is the vacuum state corresponding to the mode $\tilde{a}$ (represented as $| 0 \rangle_{a}$), $| \psi \rangle_s$ is known as the squeezed vacuum and is typically represented by $| \xi \rangle$. The original vacuum state $| 0 \rangle_{a}$ has the property that the two quadratures $\tilde{X}_{1,2}$, which do not commute and are defined by:
\begin{align}
\tilde{X}_{1} = & \frac{1}{2} \left( \tilde{a} + \tilde{a}^\dagger \right), \\
 \tilde{X}_{2} = & \frac{1}{2i} \left( \tilde{a} - \tilde{a}^\dagger \right),
\end{align}  
exhibit equal quantum fluctuations. These quadratures typically represent physically relevant quantities e.g. electric and magnetic fields for optical modes. In the squeezed state $| \xi \rangle$, the two quadratures have unequal quantum fluctuations which is seen as the squeezing of the quantum noise in one quadrature at the expense of an increase in the other. In particular with $\theta = 0$,
\begin{align}
\left\langle (\Delta \tilde{X}_{1,2})^2 \right\rangle = & \frac{1}{4} e^{\mp 2 r} ,
\end{align}
where $\Delta \tilde{X} \equiv \tilde{X} - \langle \tilde{X} \rangle$. The squeezed state additionally has several interesting properties due to its non-classical nature signified by the non-positive value of its P-function over part of the phase space~\cite{Gerry2004}. 

In a similar fashion, two-mode squeeze operator $\tilde{S}_{2} (\xi)$ and vacuum $| \xi \rangle_{2}$ may be defined in the space of the two modes represented by the annihilation operators $\tilde{a}$ and $\tilde{b}$~\cite{Gerry2004}:
\begin{align}
\tilde{S}_{2}(\xi) = & \exp \left( \xi^* \tilde{a} \tilde{b} - \xi \tilde{a}^\dagger \tilde{b}^\dagger  \right), \\
| \xi \rangle_{2} = & \tilde{S}_{2}(\xi) | 0 \rangle_{a,b}.
\end{align}
For two-mode squeezing, the relevant quadratures exhibiting unequal vacuum fluctuations involve operators for both modes and do not, in general, have a simple physical interpretation. However, the two-mode squeezed state is also non-classical with interesting properties including entanglement between the two modes~\cite{Gerry2004}. Employing Baker-Hausdorff lemma:
\begin{align}
e^{\tilde{A}} \ \tilde{B} \ e^{-\tilde{A}} = & \ \tilde{B} + \frac{1}{1!} \left[ \tilde{A}, \tilde{B}  \right] + \frac{1}{2!} \left[ \tilde{A}, \left[ \tilde{A}, \tilde{B}  \right] \right] + ... \ , 
\end{align}
we obtain the relations:
\begin{align}
\tilde{S}(\xi) \tilde{a} \tilde{S}^\dagger (\xi) = & \tilde{a} \cosh r + \tilde{a}^\dagger e^{i \theta} \sinh r, \label{1modesq} \\
\tilde{S}_{2}(\xi) \tilde{a} \tilde{S}^\dagger_{2} (\xi) = & \tilde{a} \cosh r + \tilde{b}^\dagger e^{i \theta} \sinh r, \label{2modesq}
\end{align}
which will be useful at a later stage.

Now we consider the relation between the vacua corresponding to the two kinds of excitations under consideration. The s-magnon vacuum, denoted by $| 0 \rangle_{\beta}$, is defined by:
\begin{align}\label{vacdef}
\tilde{\beta}_{\pmb{q}} \ | 0 \rangle_{\beta} =  (u_{\pmb{q}} \tilde{b}_{\pmb{q}} -  v_{\pmb{q}}^*  \tilde{b}_{-\pmb{q}}^\dagger) \ | 0 \rangle_{\beta} & = 0,
\end{align}
for all $\pmb{q}$. We first consider $\pmb{q} = \pmb{0}$ yielding:
\begin{align}
(u_{\pmb{0}} \tilde{b}_{\pmb{0}} -  v_{\pmb{0}}  \tilde{b}_{\pmb{0}}^\dagger) \ | 0 \rangle_{\beta_{\pmb{0}}} & = 0,
\end{align}
where we have taken into account that $v_{\pmb{0}}$ is real. Employing Eq. (\ref{1modesq}) and identifying $\xi_{\pmb{0}} = r_{\pmb{0}} \exp(i \theta_{\pmb{0}})$, $\cosh r_{\pmb{0}} = u_{\pmb{0}}$ and $\sinh r_{\pmb{0}} \exp(i \theta_{\pmb{0}}) = - v_{\pmb{0}}$, the equation above can be written as:
\begin{align}
 \tilde{S}(\xi_{\pmb{0}}) \tilde{b}_{\pmb{0}} \tilde{S}^\dagger (\xi_{\pmb{0}}) \ | 0 \rangle_{\beta_{\pmb{0}}} & = 0,
\end{align}
whence we obtain:
\begin{align}
\tilde{S}^\dagger (\xi_{\pmb{0}}) \ | 0 \rangle_{\beta_{\pmb{0}}} & = | 0 \rangle_{b_{\pmb{0}}} \quad \mathrm{or} \quad | 0 \rangle_{\beta_{\pmb{0}}}  = \tilde{S} (\xi_{\pmb{0}}) \ | 0 \rangle_{b_{\pmb{0}}},
\end{align}
demonstrating that the $\beta_{\pmb{0}}$ excitation is obtained by squeezing the $b_{\pmb{0}}$ excitation. The relation obtained above is complementary to the demonstration of the $\pmb{q} = \pmb{0}$ mode squeezing via evaluation of the vacuum fluctuation of the net magnetic moment x and y components presented in the main text. In an analogous fashion, using Eqs. (\ref{2modesq}) and (\ref{vacdef}), the squeezing of $\pmb{q} \neq \pmb{0}$ modes can be demonstrated as has already been discussed in the main text.

\section{Solution to equations of motion}

In this section, we give a relatively detailed derivation of the dynamical equation [Eq. (11) in the main text] for the coherently driven $\pmb{q} = \pmb{0}$ mode starting from the total Hamiltonian [Eq. (2) in the main text]. The calculation of other relevant quantities, such as current and noise, follows an analogous mathematical treatment. Since all operators of interest can be expressed in terms of the eigenmode creation and annihilation operators, the time evolution of the latter gives a complete description of the system. The Heisenberg equations of motion read:
\begin{align}
\dot{\tilde{c}}_{\pmb{k} +} = & \ \frac{1}{i \hbar} \left[ \tilde{c}_{\pmb{k} +} , \tilde{\mathcal{H}}  \right] \ = \ - i \omega_{\pmb{k}} \tilde{c}_{\pmb{k} +} - i  \sum_{\pmb{k}_2 \pmb{q}} W_{\pmb{k} \pmb{k}_2 \pmb{q}} \ \tilde{c}_{\pmb{k}_2 -} \tilde{b}_{\pmb{q}}, \label{eomckplus} \\
\dot{\tilde{c}}_{\pmb{k} -}   = & - i \omega_{\pmb{k}} \tilde{c}_{\pmb{k} -} - i  \sum_{\pmb{k}_1 \pmb{q}} W_{\pmb{k}_1 \pmb{k} \pmb{q}}^* \ \tilde{c}_{\pmb{k}_1 +} \tilde{b}_{\pmb{q}}^\dagger, \label{eomckminus} \\
\dot{\tilde{\beta}}_{\pmb{q}}   = & - i \omega_{\pmb{q}} \tilde{\beta}_{\pmb{q}} - i \sum_{\pmb{k}_1 \pmb{k}_2}  \left( u_{\pmb{q}} W_{\pmb{k}_1 \pmb{k}_2 \pmb{q}}^* \ \tilde{c}_{\pmb{k}_2 - }^\dagger \tilde{c}_{\pmb{k}_1 +} + v_{\pmb{q}} W_{\pmb{k}_1 \pmb{k}_2 \pmb{q}} \ \tilde{c}_{\pmb{k}_1 + }^\dagger \tilde{c}_{\pmb{k}_2 -} \right) \nonumber \\ 
  & + i \frac{\mu_0 h_0 B}{\hbar} \cos \omega t ~ \delta_{\pmb{q},\pmb{0}}. \label{eombetaq}
\end{align}
We aim to obtain solution to these equations perturbatively up to the second order in the interfacial exchange parameter $\mathcal{J}$ [Eq. (6) in the main text], and hence $W_{\pmb{k}_1 \pmb{k}_2 \pmb{q}}$. To this end, we use the method employed by Gardiner and Collet~\cite{Gardiner1985} in deriving the input-output formalism~\cite{Walls2008} for quantum optical fields. This method entails the following procedure. Until a certain initial time $t_0$, F and N exist in thermal equilibrium without any mutual interaction or the driving field, such that the density matrix of the combined system is the outer-product of the F and N equilibrium density matrices, i.e. $\rho = \rho_{\mathrm{F}}^{\mathrm{eq}} \otimes \rho_{\mathrm{N}}^{\mathrm{eq}}$. At $t = t_0$, the F and N interaction ($\tilde{\mathcal{H}}_{\mathrm{int}}$) and the microwave drive ($\tilde{\mathcal{H}}_{\mathrm{drive}}$) are turned on. In the Heisenberg picture, the density matrix for the system stays the same while the operators evolve with time and get entangled. The steady state dynamics is obtained by taking the limit $t_0 \to - \infty$ in the end. Within this prescription, the general solution to equation (\ref{eomckplus}) for $t > t_0$ may be written as~\cite{Gardiner1985}:
\begin{align}\label{ckplusexp}
\tilde{c}_{\pmb{k} + } (t) = & e^{- i \omega_{\pmb{k}} (t - t_0)} \tilde{c}_{\pmb{k} +}(t_0) - i \sum_{\pmb{k}_2 \pmb{q}} W_{\pmb{k} \pmb{k}_2 \pmb{q}} \ \int_{t_0}^{t} e^{- i \omega_{\pmb{k}}(t - t^\prime)} \  \tilde{c}_{\pmb{k}_2 -}(t^\prime) \tilde{b}_{\pmb{q}}(t^\prime)  dt^\prime,  
\end{align}
where $ \tilde{c}_{\pmb{k} +}(t_0)$ is the initial value of the operator. In the equation above, the first term represents the unperturbed solution while the second term gives the effect of exchange interaction $\tilde{\mathcal{H}}_{\mathrm{int}}$. A similar expression follows for $\tilde{c}_{\pmb{k} - } (t)$ using equation (\ref{eomckminus}).

Since the microwave drives the $\pmb{q} = \pmb{0}$ mode coherently, represented by the last term on the right hand side of the {\em linear} dynamical equation [(\ref{eombetaq})] for $\tilde{\beta}_{\pmb{q}}$, we may express $\tilde{\beta}_{\pmb{0}} = \beta + (\tilde{\beta}_{\pmb{0}} - \beta)$ as the sum over the coherent part given by a c-number $\beta = \langle \tilde{\beta}_{\pmb{0}} \rangle$ and the incoherent part $\tilde{\beta}_{\pmb{0}} - \beta$. The dynamical equation for $\beta$ is obtained by taking the expectation value on both sides of equation (\ref{eombetaq}) for $\pmb{q} = \pmb{0}$:
\begin{align}\label{betadyn1}
\dot{\beta}  = & - i \omega_{\pmb{0}} \beta - i \sum_{\pmb{k}_1 \pmb{k}_2}  \left( u_{\pmb{0}} W_{\pmb{k}_1 \pmb{k}_2 \pmb{0}}^* \ Y_{\pmb{k}_1 \pmb{k}_2} + v_{\pmb{0}} W_{\pmb{k}_1 \pmb{k}_2 \pmb{0}} \  Y_{\pmb{k}_1 \pmb{k}_2}^* \right) + i \frac{\mu_0 h_0 B}{\hbar} \cos \omega t,
\end{align}
with $Y_{\pmb{k}_1 \pmb{k_2}} = \langle \tilde{c}_{\pmb{k}_2 - }^\dagger \tilde{c}_{\pmb{k}_1 +} \rangle$. Employing equation (\ref{ckplusexp}) and analogous expressions for $\tilde{c}_{\pmb{k} - } (t)$ and $\tilde{\beta}_{\pmb{q}}(t)$, retaining terms up to the second order in $\mathcal{J}$, we obtain:
\begin{align}\label{yk1k2}
Y_{\pmb{k}_1 \pmb{k_2}} = & i \pi W_{\pmb{k}_1 \pmb{k}_2 \pmb{0}} \ (n_{\pmb{k}_1} - n_{\pmb{k}_2}) \ \left[ u_{\pmb{0}} \beta \delta(\omega_{\pmb{k}_1} - \omega_{\pmb{k}_2} - \omega) + v_{\pmb{0}} \beta^* \delta(\omega_{\pmb{k}_1} - \omega_{\pmb{k}_2} + \omega) \right],
\end{align}
with $n_{\pmb{k}} = \langle \tilde{c}_{\pmb{k}}^\dagger(t_0) \tilde{c}_{\pmb{k}}(t_0) \rangle = f(\hbar \omega_{\pmb{k}} - \mu)$, where $f(\epsilon) = 1/[\exp(\epsilon/k_B T) + 1]$ is the Fermi function, $\mu$ is the chemical potential in N, $k_B$ is the Boltzmann constant, and $T$ is the system temperature. Employing equation (\ref{yk1k2}), equation (\ref{betadyn1}) simplifies to the desired result [Eq. (11) in the main text]:
\begin{align}\label{betadyn2}
\dot{\beta}  = & - i \omega_{\pmb{0}} \beta - (u_{\pmb{0}}^2 + v_{\pmb{0}}^2) \Gamma_{\mathrm{N}} \beta + 2 u_{\pmb{0}} v_{\pmb{0}} \Gamma_{\mathrm{N}} \beta^* + i \frac{\mu_0 h_0 B}{\hbar} \cos \omega t,
\end{align}
where $\Gamma_{\mathrm{N}}$ is defined by:
\begin{align}\label{gamman}
\Gamma_{\mathrm{N}} \equiv \Gamma_{\mathrm{N}}(\omega) = & \sum_{\pmb{k}_1,\pmb{k}_2} \pi |W_{\pmb{k}_1 \pmb{k}_2 \pmb{0}}|^2 (n_{\pmb{k}_2} - n_{\pmb{k}_1}) \delta(\omega_{\pmb{k}_1} - \omega_{\pmb{k}_2} - \omega).
\end{align}
In writing equation (\ref{betadyn2}), we have employed the relation $\Gamma_{\mathrm{N}}(-\omega) = - \Gamma_{\mathrm{N}}(\omega)$. We now make two simplifying assumptions: (i) $|W_{\pmb{k}_1 \pmb{k}_2 \pmb{0}}|^2 \equiv |W_{\mu,\pmb{0}}|^2$, i.e. $W_{\pmb{k}_1 \pmb{k}_2 \pmb{0}}$ only depends on the magnitudes of $\pmb{k}_{1,2}$, and thus on the chemical potential in N, and (ii) the electronic density of states per unit volume in N - $g(\epsilon)$ - does not vary considerably over energy scales $k_B T$ and $\hbar \omega$ around $\epsilon = \mu$. With these assumptions, equation (\ref{gamman}) leads to the simplified expression $\Gamma_{\mathrm{N}} = \alpha^\prime \omega$, with $\alpha^\prime = \pi |W_{\mu,\pmb{0}}|^2 V_{\mathrm{N}}^2 \hbar^2 g^2(\mu)$.

\end{document}